\documentstyle[11pt,aaspp4,psfig]{article}
\newcommand{\gtsima}{$\; \buildrel > \over \sim \;$}
\newcommand{\ltsima}{$\; \buildrel < \over \sim \;$}
\newcommand{\simgt}{\lower.5ex\hbox{\gtsima}}
\newcommand{\simlt}{\lower.5ex\hbox{\ltsima}}

\newcommand{\ns}{log$N$--log$S$~}
\newcommand{\unit}{erg\,cm$^{-2}$\,s$^{-1}$}
\newcommand{\A}{{\scriptscriptstyle A}}
\newcommand{\B}{{\scriptscriptstyle B}}
\newcommand{\dL}{{d_{\scriptscriptstyle L}}}
\newcommand{\CMB}{{\rm\scriptscriptstyle CMB}}
\newcommand{\gas}{{\rm\scriptscriptstyle gas}}
\newcommand{\bol}{{\rm\scriptscriptstyle bol}}
\newcommand{\band}{{\rm\scriptscriptstyle band}}
\begin{document}
\begin{minipage}[c]{3cm}
  \psfig{figure=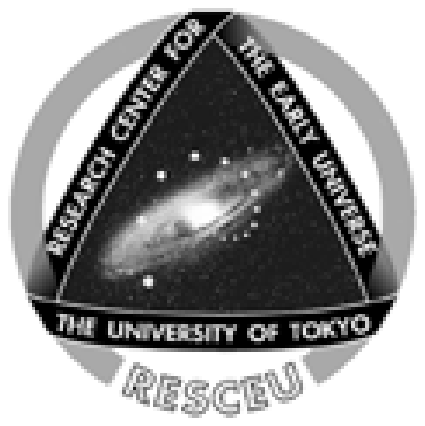,height=3cm}
\end{minipage}
\begin{minipage}[c]{9cm}
\begin{centering}
{
\vskip 0.1in
{\large \sf 
THE UNIVERSITY OF TOKYO\\
\vskip 0.1in
Research Center for the Early Universe}\\
}
\end{centering}
\end{minipage}
\begin{minipage}[c]{3cm}
\vspace{1.5cm}
RESCEU-29/97\\
UTAP-266/97
\end{minipage}\\
\vspace{0.5cm}
\title{ Cosmological Implications of Number Counts of Clusters of
  Galaxies: log$N$--log$S$ in X-Ray and Submm Bands}
\author{Tetsu {\sc Kitayama},$^{1}$
Shin {\sc Sasaki},$^{2}$
and Yasushi {\sc Suto}$^{1,3}$
\\[12pt]
$^{1}$ {\it Department of Physics, The University of Tokyo,
Tokyo 113}\\
{\it E-mail(TK): kitayama@utaphp2.phys.s.u-tokyo.ac.jp}\\
$^{2}$ {\it Department of Physics, Tokyo Metropolitan University, Hachioji,
Tokyo 192-03}\\
$^{3}$ {\it Research Center for the Early Universe,
School of Science, The University of Tokyo, Tokyo 113} \\
}
\baselineskip=14pt
\begin{abstract}
  We compute the number counts of clusters of galaxies, the
  log$N$--log$S$ relation, in several X-ray and submm bands on the
  basis of the Press--Schechter theory. We pay particular attention 
  to a set of theoretical models which well reproduce the {\it ROSAT}
  0.5-2 keV band log$N$--log$S$, and explore possibilities to further
  constrain the models from future observations with {\it ASCA} and/or
  at submm bands. The latter is closely related to the European {\it
    PLANCK} mission and the Japanese Large Millimeter and
  Submillimeter Array ({\it LMSA}) project. We exhibit that one can
  break the degeneracy in an acceptable parameter region on the
  $\Omega_0 - \sigma_8$ plane by combining the {\it ROSAT}
  log$N$--log$S$ and the submm number counts. Models which reproduce
  the {\it ROSAT} band log$N$--log$S$ will have $N(>S) \sim (150-300)
  (S/10^{-12}\mbox{erg~cm$^{-2}$~s$^{-1}$})^{-1.3}$ str$^{-1}$ at $S
  \simgt 10^{-12}$erg~cm$^{-2}$~s$^{-1}$ in the {\it ASCA} 2-10 keV
  band, and $N(>S_\nu) \sim (10^2-10^4) (S_\nu/100\mbox{mJy})^{-1.5}
  \mbox { str}^{-1}$ at $S_\nu \simgt 100\mbox{mJy}$ in the submm
  (0.85mm) band.  The amplitude of the log$N$--log$S$ is very
  sensitive to the model parameters in the submm band.  We also
  compute the redshift evolution of the cluster number counts and
  compare with that of the X-ray brightest Abell-type clusters. The
  results, although still preliminary, point to low density
  ($\Omega_0\sim 0.3$) universes.  The contribution of clusters to the
  X-ray and submm background radiations is shown to be insignificant
  in any model compatible with the {\it ROSAT} log$N$--log$S$.
\end{abstract}
\keywords{Cosmology --- Galaxies : clusters of --- Radio sources :
  extended --- X-rays : sources}

\vfill \centerline{\sl Publications of the Astronomical Society of
  Japan, in press}

%
\clearpage

\baselineskip=15pt
\section{Introduction}

Clusters of galaxies are among the largest virialized structures in
the universe and their importance as a cosmological probe is
well-recognized. Thus their observations have been actively carried
out in a variety of wavelengths including X-ray, optical, infrared,
and radio bands.  In the present paper, we focus on the theoretical
predictions for the number counts of clusters of galaxies, the \ns
relation, rather than more conventional statistics such as the X-ray
temperature and luminosity functions (hereafter XTF and XLF) for
several reasons; 1) temperature of X-ray clusters can be reliably
determined only for luminous ones, and thus the statistics is
inevitably limited, 2) such obtained XTF is to some extent weighted
towards relatively rich clusters, and may be biased for the luminous
species, 3) the XTF and XLF at high redshifts ($z \simgt 0.1$) are in
fact model-dependent statistics, because the translation of the
observed X-ray flux to the absolute luminosity, and of the observed
number to the comoving number density can be done only by assuming
specific values of the cosmological parameters (the density parameter,
$\Omega_0$, the dimensionless cosmological constant, $\lambda_0$, and
the Hubble constant $H_0$ in units of 100 km/s/Mpc, $h$).

On the other hand, the \ns relation is almost free from the above
problems as long as the cluster identification (or separation from
point-like sources) and the conversion of count rates to fluxes are
reliable. Recent analysis of the {\it ROSAT} Deep Cluster Survey
(RDCS, Rosati et al. 1995, 1997) and the {\it ROSAT} Brightest Cluster
Sample (BCS, Ebeling et al. 1997a,b) has determined the \ns of
clusters over almost four orders of magnitude in flux,
i.e. $S(\mbox{0.5-2.0 keV}) \sim 10^{-14}-10^{-10}$ \unit.  The number
of identified clusters in the \ns is over 200, an order of magnitude
larger than that for the commonly used XTF based on Henry \& Arnaud
(1991), and therefore the \ns data are statistically more reliable.

Kitayama \& Suto (1997, hereafter KS97) found that a set of cold dark
matter (CDM) models reproduce the above {\it ROSAT} \ns data
remarkably well over whole observed flux range, and simultaneously
agree with the observed XTF and the {\it COBE} 4 year data.
Nevertheless, there still exist some degeneracy of acceptable
cosmological parameters. In the present paper, we explore
possibilities to further constrain the models so as to break such
degeneracy and discuss their implications in the following manner.

First, we combine the \ns relations at different wavelengths, in X-ray
and submm bands. The latter is of particular significance in relation
to the future projects including the European {\it PLANCK} mission and
the Japanese Large Millimeter and Submillimeter Array ({\it LMSA})
project.  The emissions from intracluster gas in X-ray and submm bands
are originated from completely different physical mechanisms; the
former is mainly due to thermal bremsstrahlung, and the latter is due
to the inverse-Compton scattering of the cosmic microwave background
(CMB) photons, i.e. the Sunyaev \& Zel'dovich (1972, hereafter SZ)
effect.  As a result, the \ns relations in these bands show very
different parameter dependence.  We note that the submm \ns was
computed earlier by several authors (e.g., Barbosa et al. 1996;
Colafrancesco et al. 1997). Our analysis below differs from theirs in
considering several CDM models consistent with the {\it ROSAT} \ns
data, in including the relativistic correction to the SZ effect, and
in making quantitative and extensive predictions for the number counts
on the $\Omega_0 - \sigma_8$ plane.

Second, we consider the cluster number counts incorporating redshift
and/or temperature information in addition to the flux. In this way,
we are able to discuss the evolution of cluster abundances on the
basis of a cosmological model-independent and bias-free observable at
high redshifts, which is in contrast to the approaches based on the
XTF or XLF.  We demonstrate a tentative comparison of our predictions
with an observed sample from the {\it ROSAT} All Sky Survey, the X-ray
brightest Abell-type clusters (Ebeling et al. 1996).
 
Finally, we discuss the implications of our results for the X-ray
background (XRB) and the submm background radiation (SBR).  Since the
\ns relation is closely related to the background radiation in the
corresponding energy band, we may rigorously constrain the
contribution of clusters of galaxies to the XRB and SBR.

\section{Number counts of clusters of galaxies in X-ray and submm bands}

\subsection{X-ray flux from clusters}
\label{sec:xray}

Following KS97, we compute the number of clusters observed per unit
solid angle with X-ray flux greater than $S$ by
\begin{eqnarray}
  N(>S)= \int_{0}^{\infty}dz ~d_\A^2(z) \, c
  \left|{\frac{dt}{dz}}\right| \int_{S}^\infty dS ~ (1+z)^3 n_M(M,z)
  \frac{dM}{dT_{\gas}}\frac{dT_{\gas}}{dL_\band} \frac{dL_\band}{dS},
\label{eq:logns}
\end{eqnarray}
where $c$ is the speed of light, $t$ is the cosmic time, $d_\A$ is the
angular diameter distance, $T_{\gas}$ and $L_{\band}$ are respectively
the gas temperature and the band-limited absolute luminosity of
clusters, and $n_M(M,z)dM$ is the comoving number density of
virialized clusters of mass $M \sim M+dM$ at redshift $z$.

Given the observed flux $S$ in an X-ray energy band [$E_a$,$E_b$], the
source luminosity $L_\band$ at $z$ in the corresponding band
[$E_a(1+z)$,$E_b(1+z)$] is written as
\begin{equation}
  L_\band[E_a(1+z),E_b(1+z)] = 4 \pi \dL^2(z) S[E_a,E_b],
\label{eq:ls}  
\end{equation}
where $\dL = (1+z)^2 d_\A$ is the luminosity distance. Since the X-ray
luminosity of clusters of galaxies depends sensitively on the details
of the cluster gas density properties (distribution and clumpiness),
its theoretical prediction as a function of $M$ or $T_{\gas}$ is
difficult. In fact, a simple self-similar model predicts for the
bolometric luminosity, $L_\bol \propto T_{\gas}^2$, which is
inconsistent with the observed relation $L_\bol \propto
T_{\gas}^{3\sim3.5}$. Although a preheated cluster model might account
for the latter (Kaiser 1991; Evrard \& Henry 1991; Bower 1997), such
theoretical models have not yet been specified. Thus we adopt the
observed $L_\bol - T_\gas$ relation parameterized by
\begin{equation}
  L_\bol = L_{44} \left( \frac{T_{\gas}}{6{\rm keV}} 
\right)^{\alpha}
  (1+z)^\zeta ~~ 10^{44} h^{-2}{\rm ~ erg~sec}^{-1} .
\label{eq:lt}
\end{equation}
As in KS97, we take $L_{44}=2.9$, $\alpha=3.4$ and $\zeta=0$ as a
fiducial set of parameters on the basis of recent observational
indications (David et al. 1993; Ebeling et al. 1996; Ponman et
al. 1996; Mushotzky \& Scharf 1997). Then we translate
$L_\bol(T_{\gas})$ into the band-limited luminosity
$L_\band[T_{\gas},E_1,E_2]$ as
\begin{eqnarray}
  L_\band[T_{\gas},E_a(1+z),E_b(1+z)] = L_\bol (T_{\gas})\times
  f[T_{\gas},E_a(1+z),E_b(1+z)],
\label{eq:band}
\end{eqnarray}
where $f[T_{\gas},E_1,E_2]$ is the band correction factor which takes
account of metal line emissions (Masai 1984) in addition to the
thermal bremsstrahlung; the former makes significant contribution to
the soft band luminosity especially at low temperature. Throughout
this paper, we fix the abundance of intracluster gas as 0.3 times the
solar value. Equations (\ref{eq:ls}), (\ref{eq:lt}) and
(\ref{eq:band}) relate $S$ and $T_{\rm gas}$ through $L_{\rm band}$,
and are used to compute equation (\ref{eq:logns}).

Assuming that the intracluster gas is isothermal, its temperature
$T_{\gas}$ is related to the total mass $M$ by
\begin{eqnarray}
  k_\B T_{\gas} &=& \gamma {\mu m_p G M \over 3 r_{\rm vir}(M,z_f)},
  \nonumber \\ &=& 5.2\gamma (1+z_f) \left({\Delta_{\rm vir} \over
      18\pi^2}\right)^{1/3} \left({M \over 10^{15} h^{-1} M_\odot}
  \right)^{2/3} \Omega_0^{1/3} ~{\rm keV}.
\label{eq:tm}
\end{eqnarray}
where $k_\B$ is the Boltzmann constant, $G$ is the gravitational
constant, $m_p$ is the proton mass, $\mu$ is the mean molecular weight
(we adopt $\mu=0.59$), and $\gamma$ is a fudge factor of order unity
which may be calibrated from hydrodynamical simulations or
observations.  The virial radius $r_{\rm vir}(M,z_f)$ of a cluster of
mass $M$ virialized at $z_f$ is computed from $\Delta_{\rm vir}$, the
ratio of the mean cluster density to the mean density of the universe
at that epoch. We evaluate this quantity using the formulae for the
spherical collapse model presented in Kitayama \& Suto (1996b) and
assuming for simplicity that $z_f$ is equal to the epoch $z$ at which
the cluster is observed.

Finally, we compute the mass function $n_M(M,z)dM$ in equation
(\ref{eq:logns}) using the Press--Schechter theory (Press \& Schechter
1974) assuming $z=z_f$ as above.  The effect of $z_f \neq z$ is
discussed by KS97 in this context, and the more general consideration
of $z_f \neq z$ is given in Lacey \& Cole (1993), Sasaki (1994), and
Kitayama \& Suto (1996a,b).

\subsection{Submm flux from clusters due to the Sunyaev-Zel'dovich effect}

The inverse-Compton scattering of the CMB photons due to high
temperature electron gas leads to distortion of the CMB spectrum
(Sunyaev \& Zel'dovich 1972). If the electrons are non-relativistic,
the change in the CMB intensity observed at frequency $\nu_0$ at a
position angle $\vec\theta$ from the cluster center is given by 
\begin{eqnarray}
\label{eq:disznr}
\Delta I^{\rm NR}_\nu(x,\vec\theta) &=& \frac{2(k_\B
  T_{\CMB})^3}{(h_{\rm P}
  c)^2} g(x) y(\vec \theta), \\ 
\label{eq:gx}
g(x) &=& \frac{x^4e^x}{(e^x-1)^2} 
     \left(x\coth\frac{x}{2}-4 \right) ,\\
\label{eq:ytheta}
    y(\vec \theta) &= &
\int^{\infty}_{-\infty}\frac{k_\B T_{\gas}(z)}{m_{\rm e}
    c^2}\sigma_{\rm T} n_{\rm e}(\vec \theta, l) dl , 
\end{eqnarray}
where $x \equiv h_{\rm P} \nu_0/(k_\B T_{\CMB, 0})$, $T_{\CMB,
  0}=2.726$K is the present-day CMB temperature, $h_{\rm P}$ is the
Planck constant, $\sigma_{\rm T}$ is the Thomson cross section,
$m_{\rm e}$ is the electron mass, and $n_{\rm e}$ is the electron
number density. At frequencies $\nu_0 >217$GHz (or wavelengths
$<1.4$mm), $g(x)$ becomes positive and galaxy clusters become positive
sources. In what follows we consider mainly the submm band at the
observed wavelength 0.85mm ($\nu_0=350$GHz) where the emission from
clusters is fairly strong ($x=6.2$ and $g(x)=6.7$) and the ground
observations are feasible.

Then the total flux from a cluster located at redshift $z$ is
\begin{eqnarray}
&& S^{\rm NR}_\nu(x,M,z) = \int \Delta I^{\rm NR}_\nu(x,\vec\theta)
  d^2 \theta \nonumber \\ &&~~~ = 25.5\, h\, (1+z)\, g(x)\, {1+X \over 2}
  {\Omega_{\rm B} \over \Omega_0^{2/3}} \left[{d_\A(z) \over c
      H_0^{-1}}\right]^{-2} \left({\Delta_{\rm vir} \over
      18\pi^2}\right)^{1/3} \left({M \over 10^{15} h^{-1} M_\odot}
  \right)^{5/3} {\rm mJy},
\label{eq:szflux}
\end{eqnarray}
where $X$ is the hydrogen mass fraction for which we adopt $X=0.76$
hereafter, and $\Omega_{\rm B}$ is the baryon density parameter (we
adopt $\Omega_{\rm B}=0.0125h^{-2}$).  It should be noted that the
above flux corresponds to the value averaged over the entire cluster,
and the observed \ns might be somewhat different depending on the
details of the instruments specifically used in the survey. Aghanim et
al. (1997) address this issue in detail.

At $T_{\gas} \simgt 10$ keV, the electrons become relativistic and the
above expressions need to be modified. We thus apply the relativistic
correction to equation (\ref{eq:disznr}) derived by Rephaeli (1995;
see also Rephaeli \& Yankovitch 1997). At 0.85mm, this leads to 4\%,
11\% and 16\% reduction in flux at $T=3$keV, $8$keV and $12$keV,
respectively.  We find that the relativistic correction at this
wavelength is well fitted by
\begin{eqnarray}
  \Delta I^{\rm R}_\nu = \left[ 1 - 0.013 \left( \frac{T_{\rm gas}}{1
        \mbox{keV}} \right) \right] \Delta I_\nu^{\rm NR} \mbox{ ~~~
    at 0.85mm}.
\label{eq:szrel}
\end{eqnarray}
While clusters with $T_\gas \simgt 10$keV are rare and the above
correction does not make significant difference, we take it into
account for completeness in computing submm luminosity functions and
the \ns. Since the above submm flux after the relativistic correction
is explicitly expressed as a function of $M$ and $z$, the submm \ns is
computed in a similar manner to equation (\ref{eq:logns}).

\section{Number counts in X-ray and submm bands 
predicted in the cold dark matter models}

KS97 shows that CDM models with a certain range of parameters
reproduce the {\it ROSAT} \ns data as well as the XTF (Henry \& Arnaud
1991) and the {\it COBE} 4 year data (Bunn \& White 1997).  To be
definite, we hereafter consider five models with different sets of
parameters summarized in table 1, and see if one can break the
degeneracy of the models by comparing the \ns relations in different
bands.  Throughout this paper, we assume that the primordial spectral
index $n$ is equal to unity and use the fitting formulae given in
Kitayama \& Suto (1996b) for the CDM mass fluctuation spectrum on the
basis of Bardeen et al. (1986) transfer function.
 
\begin{figure}
\begin{center}
  \leavevmode\psfig{figure=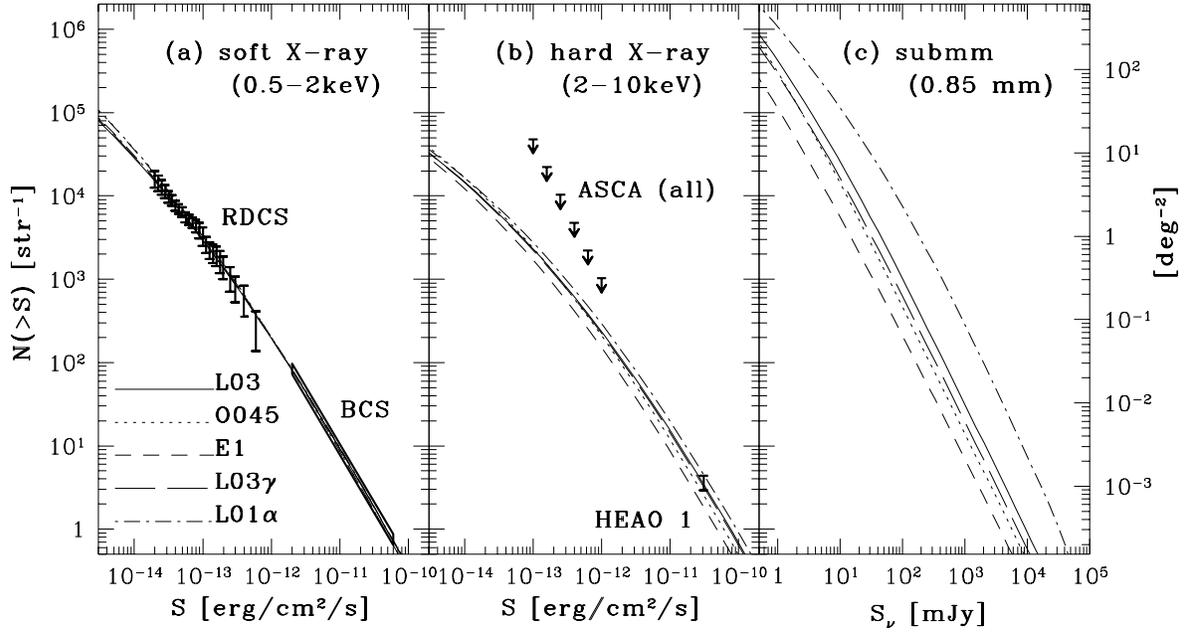,width=16cm}
\end{center}
\caption{The \ns relations of galaxy clusters for CDM models  
  in (a) the soft X-ray (0.5-2.0 keV) band, (b) the hard X-ray (2-10
  keV) band, and (c) the submm (0.85 mm) band. Lines represent the
  models listed in table 1; L03 (solid), O045 (dotted), E1 (short
  dashed), L03$\gamma$ (long dashed), and L01$\alpha$
  (dot-dashed). Also plotted in panel (a) are the 1$\sigma$ error bars
  from the RDCS (Rosati et al. 1995, 1997), and the error box from the
  BCS (Ebeling et al. 1997a,b).  In panel (b), the arrows indicate the
  \ns of all X-ray sources in the 2-10 keV band from {\it ASCA}
  (Cagnoni et al. 1997), and the error bar (1 $\sigma$) is the number
  of clusters observed by {\it HEAO 1} (Piccinotti et al. 1982).}
\label{fig:ns3} 
\end{figure}

Figure \ref{fig:ns3} shows the \ns of clusters of galaxies in the soft
X-ray (0.5-2 keV), hard X-ray (2-10 keV), and submm (0.85mm) bands. In
the soft X-ray band, all the adopted models by definition reproduce
well the observed \ns from the {\it ROSAT} Deep Cluster Survey (RDCS,
Rosati et al. 1995, 1997) and the {\it ROSAT} Brightest Cluster Sample
(BCS, Ebeling et al. 1997a,b). In the hard X-ray band, massive
clusters make slightly larger contribution to the predicted \ns than
in the soft band, and thus the lower $\Omega_0$ models with flatter
fluctuation spectra (i.e., more power at large scales) yield greater
\ns. All the models are shown to lie under the upper limit inferred
from the number of all X-ray sources observed by {\it ASCA} at 2-10
keV (Cagnoni et al. 1997), and they (except E1) are also consistent
with the number of clusters observed by {\it HEAO 1} (Piccinotti et
al. 1982) within the 1$\sigma$ errors. The predicted count in model E1
is smaller than the {\it HEAO 1} result at the $3\sigma$ level
(possible incompleteness in the {\it HEAO 1} result would simply raise
the observed data point and does not reconcile the discrepancy).

In the submm band, on the other hand, the contribution from distant
clusters becomes larger than in the X-ray bands
(eqs~[\ref{eq:ls}]--[\ref{eq:tm}] and [\ref{eq:szflux}]). Thus the
lower $\Omega_0$ models with slower evolution of fluctuations (i.e.,
more clusters at high $z$) yield greater \ns as seen in figure
\ref{fig:ns3}(c). Compared with the hard X-ray band, the degeneracy of
the models are broken in much greater extent in the submm band.
Therefore, future determination of the submm cluster counts will
enable us to constrain the models more tightly.

\begin{figure}
\begin{center}
  \leavevmode\psfig{figure=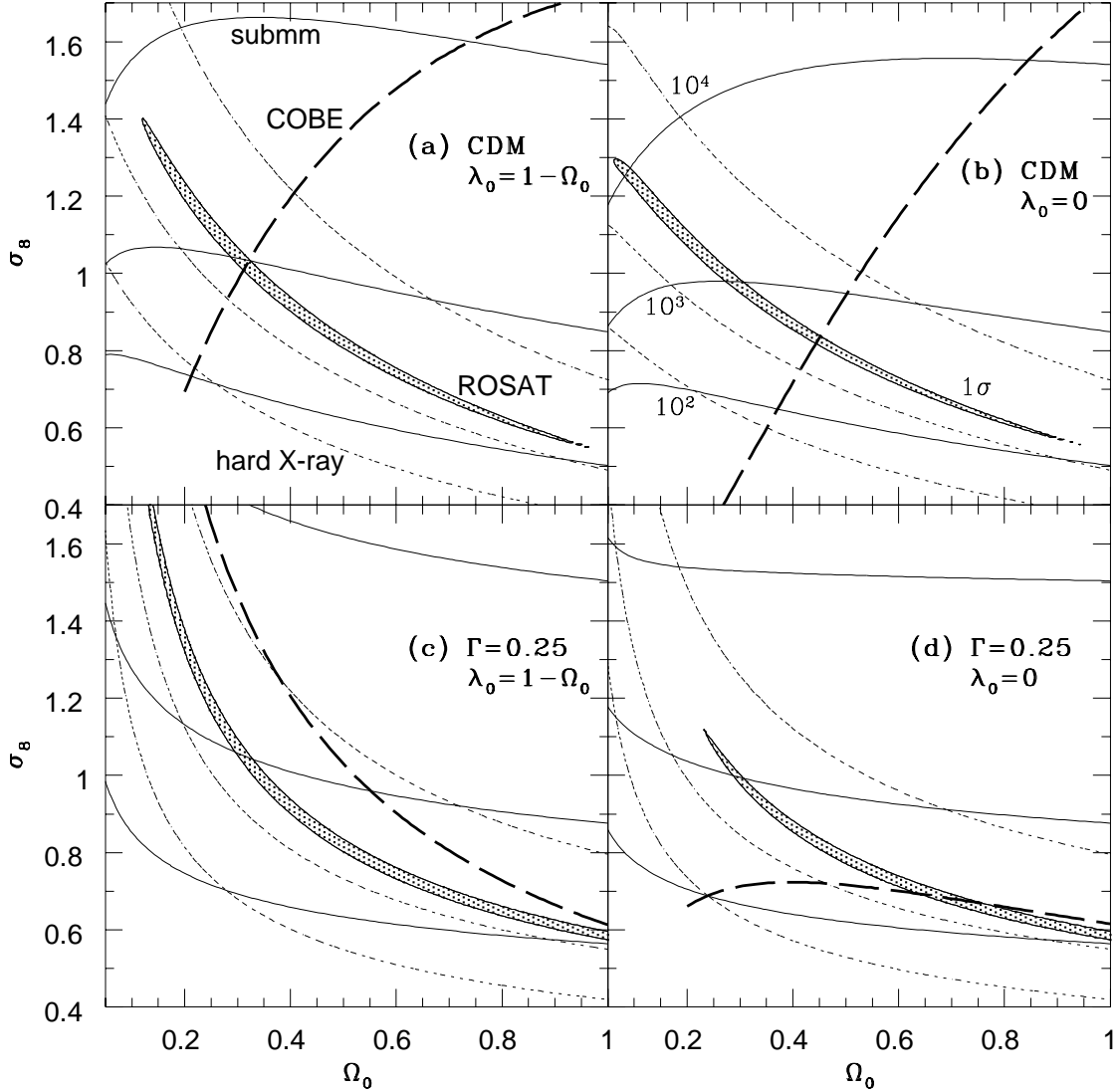,width=16cm}
\end{center}
\caption{Contour maps on the $\Omega_0$-$\sigma_8$ plane in (a) 
  spatially flat ($\lambda_0=1-\Omega_0$) CDM models, (b) open
  ($\lambda_0=0$) CDM models, (c) spatially flat CDM-like models with
  the fixed shape parameter ($\Gamma=0.25$), and (d) open CDM-like
  models with $\Gamma=0.25$.  In all cases, $h=0.7$, $\alpha=3.4$, and
  $\gamma=1.2$ are assumed. Shaded regions represent the 1$\sigma$
  significance contours derived in KS97 from the soft X-ray (0.5-2
  keV) \ns. Dotted and solid lines indicate the contours of the number
  of clusters greater than $S$ per steradian ($10^2$, $10^3$, $10^4$
  from bottom to top) with $S = 10^{-13}$ \unit in the hard X-ray
  (2-10 keV) band and with $S_\nu = 10^2$mJy in the submm (0.85 mm)
  band, respectively.  Thick dashed lines represent the {\it COBE} 4
  year result computed from the fitting formulae at
  $0.2<\Omega_0\leq1$ by Bunn \& White (1997).}
\label{fig:cont}
\end{figure}

To see this clearly, we plot in figure \ref{fig:cont} the contour maps
of the cluster \ns in different bands on the $\Omega_0-\sigma_8$
plane. For the 0.5-2 keV band, the $1-\sigma$ significance level
derived from the $\chi^2$ test between our theoretical prediction and
the observations (see KS97 for detail) is plotted, while for the 2-10
keV and submm bands, the contours of the number of clusters per
steradian ($10^2$, $10^3$, $10^4$ from bottom to top) at $S(\mbox{2-10
  keV})=10^{-13}$ \unit and $S_\nu(\mbox{0.85mm})=10^2$mJy are plotted
respectively. The $\sigma_8$ values derived from the {\it COBE} 4 year
data (Bunn \& White 1997) is also shown for reference. Panels (a) and
(b) show that the contours for the 2-10 keV band counts run almost
parallel to the $\chi^2$ contour of the 0.5-2 keV band counts. In this
sense, the future {\it ASCA} \ns data will provide an independent
consistency check of the {\it ROSAT} data.  The shape of the submm \ns
contours, on the other hand, is quite different, especially at high
$\sigma_8$, and thus should place complementary constraints on
$\Omega_0$ and $\sigma_8$.

In the conventional CDM power spectrum we have adopted, the spectral
shape vary sensitively with $\Omega_0$, $h$ and $\Omega_{\rm B}$
through the shape parameter $\Gamma=\Omega_0 h \exp(-\Omega_{\rm
  B}-\sqrt{2h}\Omega_{\rm B}/\Omega_0)$ (Sugiyama 1995). In order to
segregate the effects of changing the spectral shape, we further
consider in the lower panels of figure \ref{fig:cont} the `CDM-like'
spectrum with the fixed shape parameter $\Gamma =0.25$ inferred from
the galaxy correlation function (e.g., Peacock 1996) independent of
$\Omega_0$, $h$ and $\Omega_{\rm B}$.  The corresponding \ns contours
are similar to the conventional CDM case except at $\Omega_0 \simlt
0.2$ and $\Omega_0 \simgt 0.8$, while the {\it COBE} normalized
$\sigma_8$ is very sensitive to the changes in the spectral shape.
 
\section{Evolution of cluster abundances}

The evolutionary behavior of the cluster abundance would provide a
useful probe of the cosmological parameters as discussed by a number
of authors (e.g., Viana \& Liddle 1996; Eke, Cole, Frenk 1996; Oukbir
\& Blanchard 1997; Mathiesen \& Evrard 1997).  It is of particular
importance because the current and near future observations will
provide us with the rapidly increasing amount of information on high
redshift clusters. Thus we consider the evolution of cluster
distribution and discuss whether one can break the degeneracy among
the models inferred from the soft X-ray \ns relation (table 1).

\begin{figure}
\begin{center}
  \leavevmode\psfig{figure=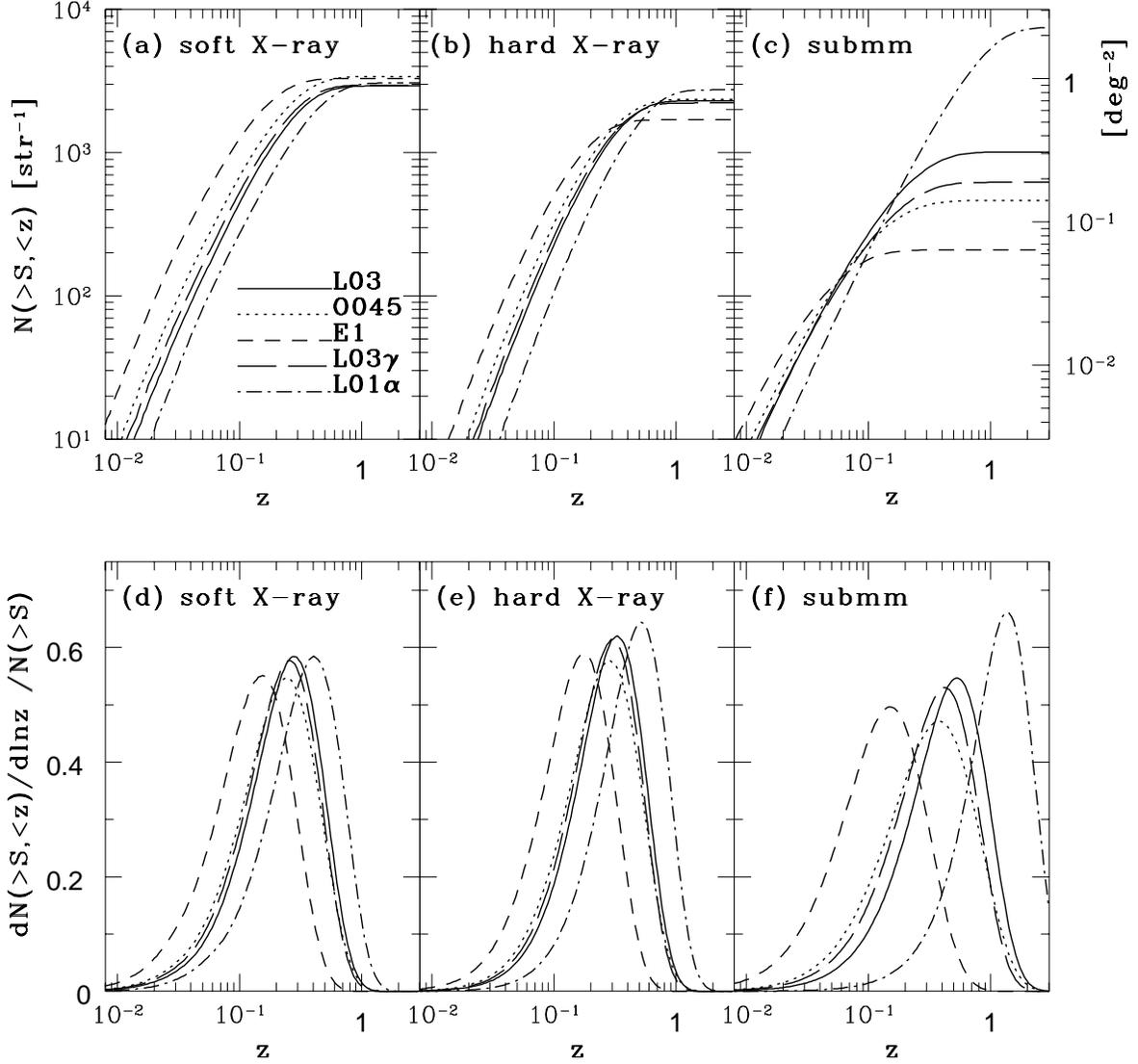,width=16cm}
\end{center}
\caption{Redshift evolution of the number of galaxy clusters. Upper 
  panels show the cumulative number $N(>\!S,<\!z)$ against $z$ in (a) the
  soft X-ray (0.5-2 keV) band with $S=10^{-13}$ \unit, (b) the hard
  X-ray (2-10 keV) band with $S=10^{-13}$ \unit, and (c) the submm
  (0.85 mm) band with $S_\nu=10^2$ mJy. Lower panels (d)--(f) are
  similar to (a)--(c) except for plotting the differential
  distribution $dN(>\!S,<\!z)/d\ln z$ normalized by $N(>S)$.}
\label{fig:nsz}
\end{figure}

Figure \ref{fig:nsz} exhibits the redshift evolution of the number of
clusters in different bands. As expected, the evolutionary behavior
strongly depends on the values of $\Omega_0$ and $\sigma_8$; the
fraction of low redshift clusters becomes larger for greater
$\Omega_0$ and smaller $\sigma_8$. It is indicated that one may be
able to distinguish among these models merely by determining the
redshifts of clusters up to $z \sim 0.2$ (see also discussion below).
 
In order to characterize the physical properties of galaxy clusters,
we have so far focused on their flux, because it is the simplest
quantity which can be determined primarily from observations.  Another
major quantity of clusters is their temperature. Although the sample
of clusters with measured temperature is still limited, an increasing
amount of temperature information is going to become available from
the {\it ASCA} (and future) observations. Furthermore, the temperature
has an advantage in that it is insensitive to the detailed
distribution of intracluster gas and much easier to model, compared to
the X-ray flux or luminosity.  This is in fact the main reason why
many of the previous analysis of cluster abundance and its evolution
are based upon the temperature rather than the X-ray luminosity
(White, Efstathiou \& Frenk 1993; Viana \& Liddle 1996; Eke et
al. 1996). However, the observed samples of clusters are usually flux
limited, which needs to be kept in mind when one compares the
theoretical predictions with the observed statistics.  In our present
framework, it is possible to incorporate explicitly the effects of
limiting fluxes in the analysis based upon the cluster temperature.

\begin{figure}
\begin{center}
  \leavevmode\psfig{figure=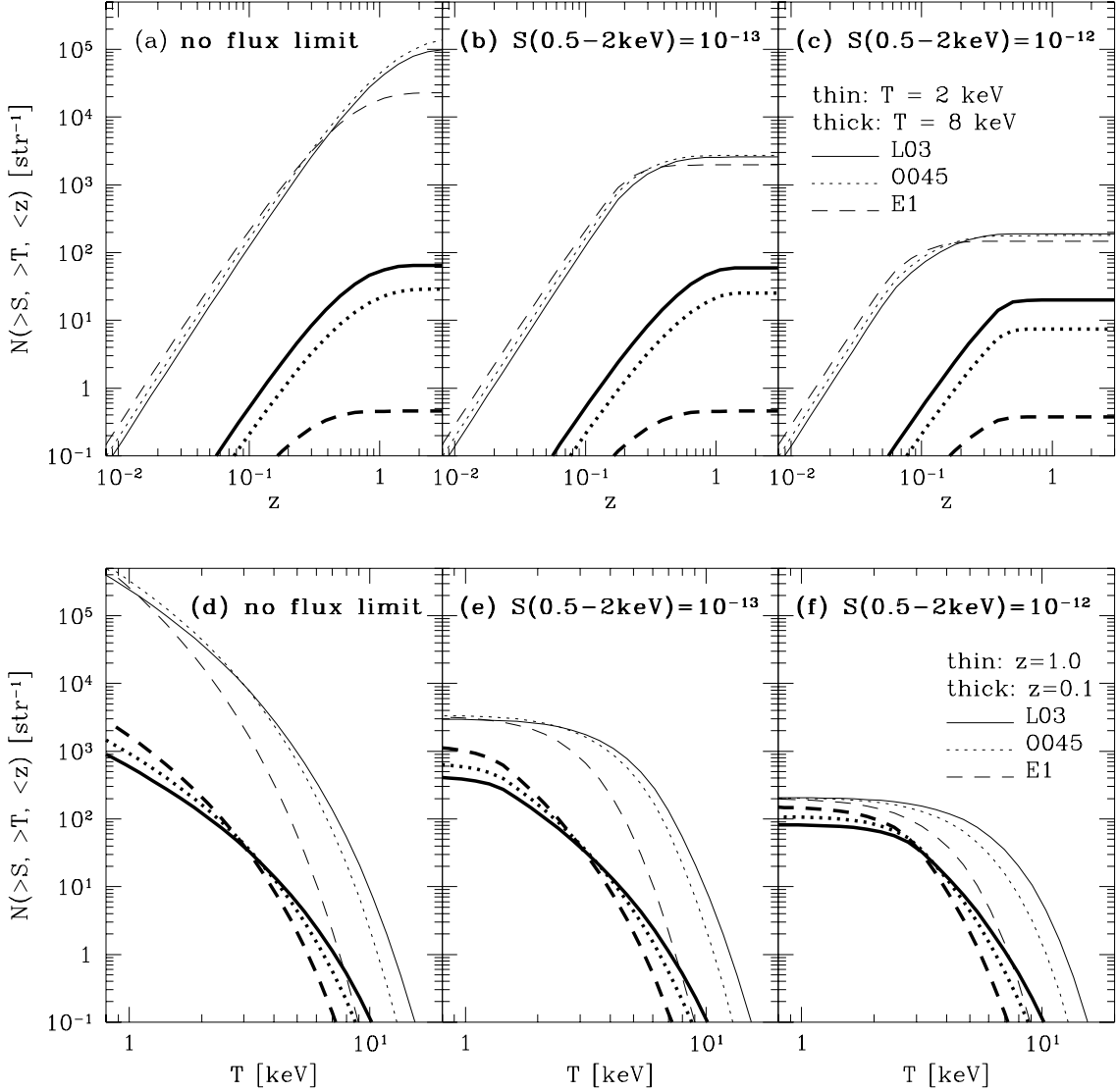,width=16cm}
\end{center}
\caption{Effects of limiting fluxes on the redshift and temperature
  distributions of galaxy clusters. Upper panels exhibit the number
  $N(>\!S,>\!T,<\!z)$ against $z$ for $T=2$keV (thin lines) and
  $T=8$keV (thick) in the cases of (a) $S=0$ (no flux limit), (b)
  $S(\mbox{0.5-2keV})=10^{-13}$ \unit, and (c)
  $S(\mbox{0.5-2keV})=10^{-12}$ \unit.  Lower panels (d)--(f) are
  similar to (a)--(c) except for plotting $N(>\!S,>\!T,<\!z)$ versus
  $T$ in the cases of $z=1$ (thin) and $z=0.1$ (thick).}
\label{fig:nstz}
\end{figure}

Figure \ref{fig:nstz} shows the cumulative distribution of clusters
with temperature greater than $T$, redshift less than $z$, and
with/without the X-ray flux limit.  It is apparent that the presence
of a flux limit affects significantly the temperature distribution and
its evolution, especially at low $T$ and high $z$. If the flux limit
is $S=10^{-12}$ \unit in the 0.5-2 keV band, for instance, the number
of clusters observed with $T>2$keV and $z<0.3$ per unit solid angle is
an order of magnitude less than what would have been expected without
any flux limit.

The above discussion has clarified that the redshift evolution (at low
$z$) and the temperature distribution of a given cluster sample with a
specific flux limit can provide a useful probe of cosmological models.
Unfortunately, we do not yet have a complete X-ray cluster sample with
redshift and temperature information. However, one may still
demonstrate the power of this approach using currently available 'best'
sample. For this purpose, we adopt tentatively the X-ray brightest
Abell-type clusters (XBACs, Ebeling et al. 1996), which is about 80 \%
complete and consists of 242 clusters with $S>5 \times 10^{-12}$ \unit
in the 0.1-2.4 keV band and $z<0.2$.  We use the temperature and
redshift data compiled in table 3 of Ebeling et al. (1996). About 92
\% and 30 \% of all the clusters listed in this table have measured
redshift and temperature, respectively. For the rest of the sample,
the table also provides the redshifts estimated from the magnitude of
the 10th-ranked cluster galaxy, and the temperatures estimated from
the empirical $L-T$ relation.  Keeping in mind the incompleteness of
the sample and uncertainties especially in the estimated temperature
data, we simply intend to perform a crude comparison with our
predictions.

\begin{figure}
\begin{center}
   \leavevmode\psfig{figure=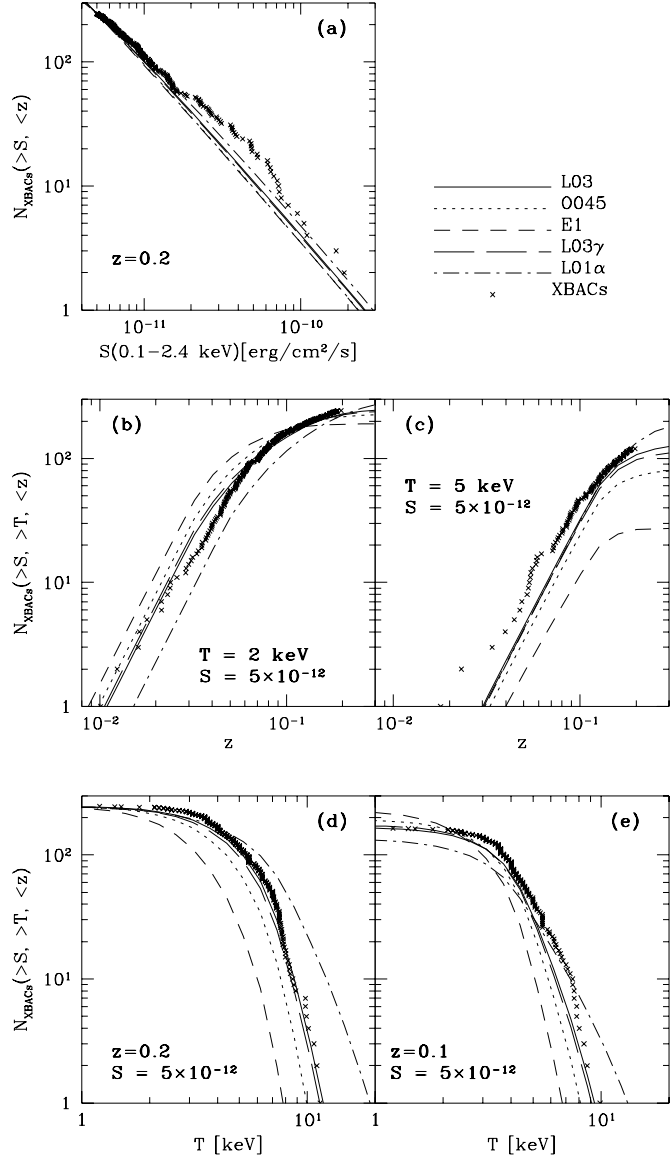,width=16cm}
\end{center}
\caption{Tentative comparison with the XBACs (Ebeling
  et al. 1996). Upper panel (a) shows the \ns in the 0.1-2.4 keV band
  with the redshift limit of $z=0.2$ (models L03 and O045 almost
  overlap with L03$\gamma$ and E1 respectively). Middle panels exhibit
  $N(>\!S,>\!T,<\!z)$ versus $z$ for $S(\mbox{0.1-2.4 keV})=5 \times
  10^{-12}$ \unit in the cases of (b) $T=2$keV and (c) $T=5$keV. Lower
  panels plot $N(>\!S,>\!T,<\!z)$ versus $T$ for $S(\mbox{0.1-2.4
    keV})=5 \times 10^{-12}$ \unit in the cases of (d) $z=0.2$ and (e)
  $z=0.1$.  All the model predictions are normalized to reproduce the
  total number of the XBACs at its flux limit $S(\mbox{0.1-2.4 keV})=5
  \times 10^{-12}$ \unit and redshift limit $z=0.2$.}
\label{fig:xbacs}
\end{figure}

The results are plotted in figure \ref{fig:xbacs}. The sky coverage of
the XBACs is hard to quantify mainly due to the uncertain volume
incompleteness of the underlying optical catalogue as noted by Ebeling
et al. (1996).  In figure \ref{fig:xbacs}, therefore, we simply plot
the real numbers of the XBACs and normalize all our model predictions
to match the total number of the XBACs at its flux limit
$S(\mbox{0.1-2.4 keV})=5 \times 10^{-12}$ \unit and redshift limit
$z=0.2$.  In general, the model predictions are shown to be capable of
reproducing well the shape and amplitude of the observed distributions
$N(>\!S,>\!T,<\!z)$ even when $T$ and $z$ are varied.  Taking into
account the incompleteness of the observed data and large statistical
fluctuations at low numbers, the agreements with models L03 and
L03$\gamma$ (both has $\Omega_0=0.3$) are rather remarkable. Since the
shapes and amplitudes of the predicted curves in these figures are
primarily determined by the value of $\Omega_0$, this result provides
a further indication for low $\Omega_0$ universe.

\section{Contribution to the X-ray and submm background radiation}

The currently observed X-ray \ns especially at faint flux end (Rosati
et al. 1997; Rosati \& Della Ceca 1997) places a model-independent
constraint on the cluster contribution to the soft XRB. From the
observed \ns relation in figure \ref{fig:ns3}(a), the cluster
contribution to the XRB is estimated roughly as
\begin{eqnarray}
  && \hspace*{-0.5cm} I^{\rm cl}(\mbox{0.5-2 keV}) = \int_{S_{\rm
      min}}^\infty S \left|\frac{dN}{dS}\right| dS \nonumber \\ &&
  \hspace*{-0.5cm} \sim 2 \times 10^{-10} \left(\ln
    \frac{S_{12}}{S_{\rm min}} + 4.3\right) \mbox{ \unit str$^{-1}$},
\label{eq:estxrb} 
\end{eqnarray}
where $S_{12} \equiv 10^{-12}$ \unit, and we have fitted the 0.5-2 keV
\ns by a broken power-law: $N(>S)\propto S^{-1}$ at $S > S_{12}$ and
$N(>S)\propto S^{-1.3}$ at $S < S_{12}$. The above estimate depends
only weakly on the faint end flux $S_{\rm min}$, and taking for
instance $S_{\rm min} = 10^{-15}$ \unit yields $I_{\rm cl}(\mbox{0.5-2
  keV}) \sim 2 \times 10^{-9}$\unit str$^{-1}$, which is less than 10
\% of the observed XRB (Gendreau et al. 1995; Suto et al. 1996).

To be more specific, we also compute numerically the contribution of
clusters to the XRB.  The XRB intensity $I^{\rm cl}_\nu(E_0)$ from
clusters at energy $E_0$, as seen by an observer at present, is given
by
\begin{eqnarray}
\label{xrb-i}
I^{\rm cl}_{\nu}(E_0) = \frac{c}{4 \pi H_0} \int_0^{\infty} dz
\frac{J_\nu[E_0(1+z),z]} {(1+z) \sqrt{\Omega_0 (1+z)^3 - K (1+z)^2 +
    \lambda_0}},
\end{eqnarray}
where $J_\nu(E,z)$ is the comoving space-averaged volume emissivity at
redshift $z$. We estimate this quantity by 
\begin{equation}
J_\nu(E,z) = \int_0^{\infty} L_\nu(E,M,z)\, n_M(M,z)\,  dM,
\end{equation}
where the luminosity $L_\nu(E,M,z)$ of a cluster of mass $M$ at $z$ is
computed in a similar manner to Section \ref{sec:xray}.  It should be
noted that the observed XRB spectrum is not derived from all sky
surveys (in particular, at lower energy band), but from small regions
in the sky where there are very few known bright X-ray sources. In
order to compare with such observations, therefore, we also need to
omit bright sources in our theoretical predictions by equation
(\ref{xrb-i}). For this purpose, we only consider contributions from
clusters with flux below the critical value $S_{\rm crit}(\mbox{0.5-2
  keV}) = 10^{-13} ~\mbox{\unit}$, which roughly corresponds to the
flux of the brightest X-ray source observed in the Lockman Hole by
{\it ROSAT} (Hasinger et al. 1993).

\begin{figure}
\begin{center}
   \leavevmode\psfig{figure=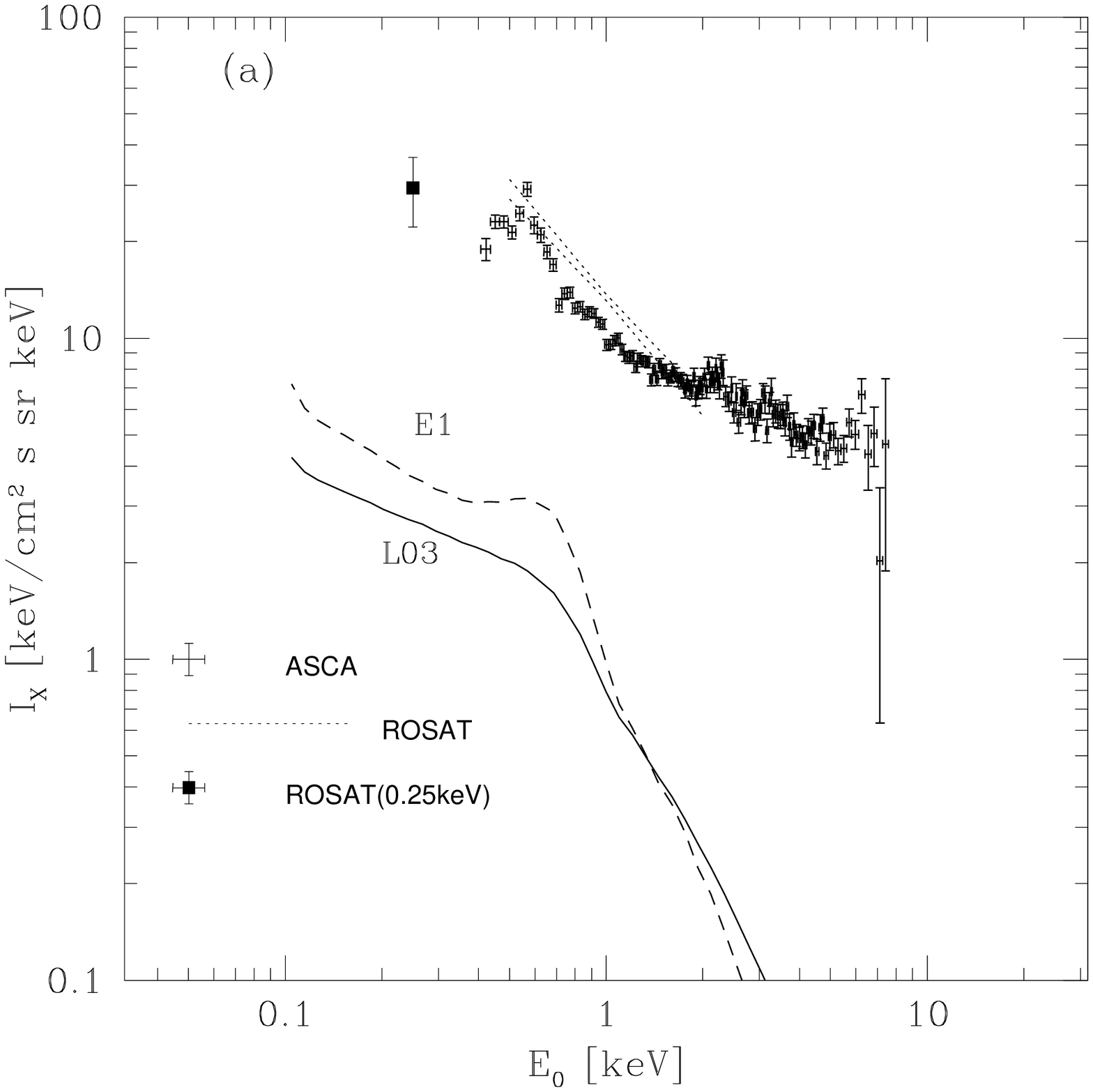,height=7.5cm}
\end{center}
\begin{center}
  \leavevmode\psfig{figure=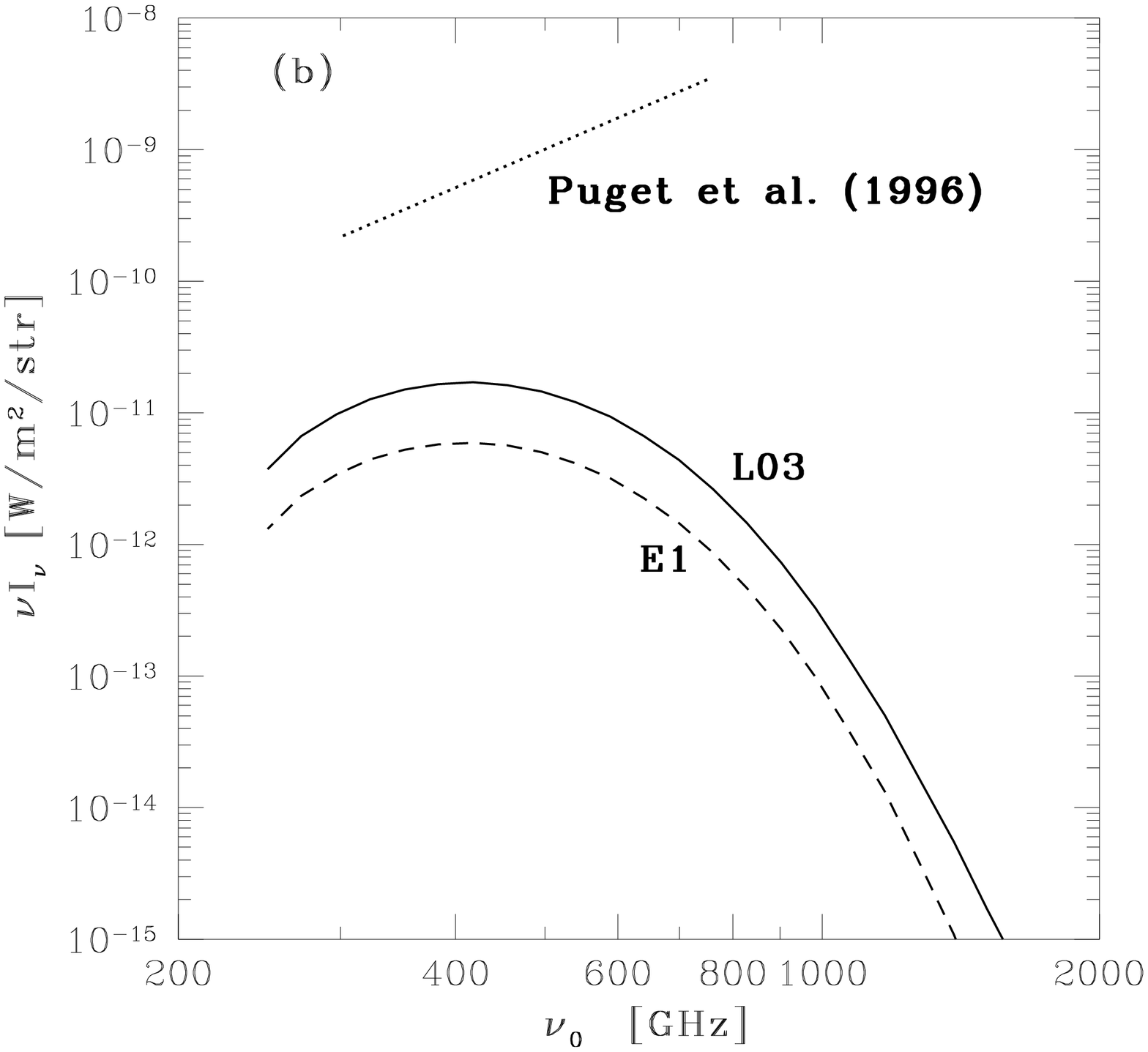,height=7.8cm}
\end{center}
\caption{Contribution of clusters of galaxies to (a) the XRB, and (b) the 
  SBR, in models L03 (solid) and E1 (dashed).  Also shown are the
  observed XRB data of {\it ASCA} (Gendreau et al. 1995), {\it ROSAT}
  (Hasinger 1992), {\it ROSAT} at 0.25 keV (Barber et al. 1996), and
  the tentative detection of the cosmic far-infrared background by
  {\it COBE} (Puget et al. 1996). }
\label{fig:bg}
\end{figure}

Figure \ref{fig:bg}a exhibits the XRB intensity from clusters of
galaxies in the cases of models L03 and E1, both of which reproduce
the observed \ns relation in the soft X-ray (0.5-2 keV) band. In these
models, clusters contribute to the XRB only less than 20 \% at $E_0
\simlt 2 {\rm keV}$, which is consistent with the rough estimate
described above. Thus, clusters of galaxies cannot be the major
sources for the XRB even in soft X-ray bands (Evrard \& Henry 1991;
Oukbir, Bartlett, Blanchard 1997).  As is clear from the above
analysis, this conclusion is based on the observed \ns relation, and
almost independent of the assumed cosmological parameters. In fact, if
one adopts a theoretical $L_\bol - T$ relation inferred from the
self-similar assumption (Kaiser 1986) which yields $\alpha=2$ in
equation (\ref{eq:lt}), there exist some models in which clusters
account for the entire soft XRB (Blanchard et al. 1992; Kitayama \&
Suto 1996a). Such models, however, are simply in conflict with both
the observed $L_\bol - T$ relation ($\alpha=3\sim3.5$) and the {\it
  ROSAT} \ns.

A similar analysis can be performed for the SBR. Since we do not yet
have the observed \ns in the submm band, we fit the predicted \ns of
model L03 and obtain 
\begin{eqnarray}
  \hspace*{-5mm} \nu I^{\rm cl}_\nu (\mbox{0.85mm}) \sim 10^{-12}
  \left(\frac{S_{\nu, {\rm min}}}{100 \mbox{mJy}}\right)^{-0.5}
  \mbox{W~m$^{-2}$~str$^{-1}$}, 
\end{eqnarray}
where we have assumed a single power law with $N(>S_\nu) \propto
S_\nu^{-1.5}$. Taking $S_{\nu, {\rm min}} = 1$mJy gives $\nu
I^{cl}_\nu (\mbox{0.85mm}) \sim 10^{-11}$W~m$^{-2}$~str$^{-1}$, which
is about 3\% of the detected SBR by Puget et al. (1996). Results from
the numerical integration shown in figure~\ref{fig:bg}b confirm that
the clusters of galaxies contribute only less than $\sim 5$\% of the
SBR.

\section{Conclusions}

We have presented several cosmological implications of the number
counts of clusters of galaxies. We have paid particular attention to
the theoretical models which are in good agreement with the {\it
  ROSAT} \ns in the soft X-ray (0.5-2 keV) band, and explored
possibilities to further constrain the models from future observations
in the {\it ASCA} hard X-ray and submm bands.

In the submm band ($0.85$mm), models which reproduce the {\it ROSAT}
\ns predict $N(>S_\nu) \sim (10^2-10^4) (S_\nu/100\mbox{mJy})^{-1.5}
\mbox { str}^{-1}$ at $S_\nu\simgt 100\mbox{mJy}$.  We have shown that
the amplitude of the above relation depend sensitively on $\Omega_0$
and $\sigma_8$, and in a substantially different manner from the {\it
  ROSAT} \ns.  Thus, combining the two can break the degeneracy in the
acceptable parameter region on the $\Omega_0 - \sigma_8$ plane. This
indicates that the future observations by the European {\it PLANCK}
mission and the Japanese {\it LMSA} project would provide powerful
probes of these parameters.

In the 2-10 keV band, the number counts show similar parameter
dependence to those in the {\it ROSAT} 0.5-2 keV band, and we predict
$N(>S) \sim 200 (S/10^{-12}\mbox{\unit})^{-1.3}$ str$^{-1}$ at
$S\simgt 10^{-12}$\unit in the {\it ASCA} 2-10 keV band.  The {\it
  ASCA} \ns would therefore provide an important cross-check for our
interpretation of the {\it ROSAT} \ns data.

The evolutionary behavior of the number counts is also important to
put additional cosmological constraints. We have exhibited that, given
a complete flux limited cluster sample with redshift and/or
temperature information, one can further constrain the cosmological
models. We have performed a tentative comparison between our
theoretical predictions and the recent compilation of the XBACs by
Ebeling et al. (1996), which is the largest sample of galaxy clusters
available to date.  While the incompleteness of the sample and
uncertainties in the temperature data still make it difficult to draw
any definite conclusions from this comparison, it is interesting to
note that our predictions reproduce well the evolutionary features of
the XBACs and that the results, although preliminary, seem to favor
low density ($\Omega_0 \sim 0.3$) universes.

The cluster \ns also provides a tight constraint on their contribution
to the background radiation in the corresponding energy band. Based on
the \ns relation observed by {\it ROSAT} in the 0.5-2 keV band, we
conclude that clusters of galaxies contribute at most $\sim 20$\% of
the total XRB and less than $\sim 5$\% of the SBR.

\bigskip 
\bigskip

\vspace{1pc} \par We are grateful to H. Ebeling and P. Rosati for
generously providing us their X-ray data and helpful comments, and to
K. Masai for making his X-ray code available to us. We also thank
A. Blanchard for valuable discussions, Y. Rephaeli for useful
correspondences on the SZ effect, and the referee G. Zamorani for
constructive comments. T.K. acknowledges support from a JSPS (Japan
Society for the Promotion of Science) fellowship (09-7408). This
research was supported in part by the Grants-in-Aid for the
Center-of-Excellence (COE) Research of the Ministry of Education,
Science, Sports and Culture of Japan (07CE2002) to RESCEU (Research
Center for the Early Universe).


%
%
\bigskip
\bigskip
\baselineskip12pt
\parskip2pt

\section*{References}
\newcommand{\re}{\par\hangindent=0.5cm\hangafter=1\noindent}

\re
Aghanim, A., De Luca, A., Bouchet, F.R., Gispert, R., Puget, J.L.\
1997, astro-ph/9705092
\re
Barber, C. R., Roberts, T. P., Warwick, R. S. \ 1996, MNRAS 282, 157
\re
Barbosa, D., Bartlett, J. G., Blanchard, A., Oukbir, J.\ 1996, A\&A 314, 13
\re
Bardeen, J. M., Bond, J. R., Kaiser, N., Szalay, A. \ 1986, ApJ 304, 15
\re
Blanchard, A., Wachter, K., Evrard, A.E., Silk, J.\ 1992, ApJ 391, 1
\re 
Bower, R. G. \ 1997, astro-ph/9701014
\re
Bunn, E. F., White, M. \ 1997, ApJ 480, 6
\re
Cagnoni, I., Della Ceca, R., Maccacaro, T. \ 1997, astro-ph/9709018 
\re 
Colafrancesco, S., Mazzotta, P., Rephaeli, Y., Vittorio, N. \ 1997, 
ApJ 479, 1 
\re
David, L. P., Slyz, A., Jones, C., Forman, W., Vrtilek, 
S. D. \ 1993, ApJ 412, 479
\re
Ebeling, H., Voges, W., B\"{o}hringer, H., Edge, A. C.,
Huchra, J. P., Briel, U. G. \ 1996, MNRAS 281, 799
\re
Ebeling, H., Edge, A. C., Fabian, A.C., Allen, S. W.,
Crawford C. S. \ 1997a, ApJ 479, L101
\re
Ebeling H., et al. \ 1997b, MNRAS submitted 
\re
Eke, V. R., Cole, S., Frenk, C. S. \ 1996, MNRAS 282, 263
\re
Evrard, A.E., Henry, J. P. \ 1991, ApJ 383, 95
\re
Gendreau, K.C. et al. \ 1995, PASJ 47, L5
\re
Hasinger, G. \ 1992, in The X-ray Background, eds. Barcons, X., 
Fabian, A. C. (Cambridge U.P.: Cambridge), 229
\re 
Hasinger, G., Burg, R., Giacconi, R., Hartner, G., Schmidt, M.,
Tr\"{u}mper, J., Zamorani, G. \ 1993, A\&A 275, 1   
\re
Henry, J. P., Arnaud, K. A. \ 1991, ApJ 372, 410
\re
Kaiser N. \ 1986, MNRAS 222, 323 
\re
Kaiser N. \ 1991, ApJ 383, 104 
\re
Kitayama, T., Suto, Y.\ 1996a, MNRAS 280, 638
\re
Kitayama, T., Suto, Y.\ 1996b, ApJ 469, 480
\re
Kitayama, T., Suto, Y.\ 1997, ApJ 490, in press (KS97)
\re
Lacey, C. G., Cole, S. \ 1993, MNRAS 262, 627
\re
Masai, K. \ 1984, Ap\&SS 98, 367
\re
Mathiesen, B., Evrard, A. E. \ 1997, astro-ph/9703176 
\re
Mushotzky, R.F., Scharf, C. A. \ 1997, ApJ 482, L13
\re
Oukbir, J., Blanchard, A. \ 1997, A\&A 317, 10
\re
Oukbir, J., Bartlett, J. G., Blanchard, A. \ 1997, A\&A 320, 365
\re
Peacock J. A. \ 1996, MNRAS 284, 885 
\re
Piccinotti, G., Mushotzky, R. F., Boldt, E. A., Holt, S. S., Marshall, 
F. E., Serlemitsos, P. J., Shafer, R. A. \ 1982, ApJ 253, 485   
\re
Ponman, T. J., Bourner, P. D. J., Ebeling, H., B\"{o}hringer, H. \ 
1996, MNRAS 283, 690
\re
Press, W. H.,  Schechter, P. \ 1974, ApJ 187, 425 (PS)
\re
Puget, J.-L., Abergel, A., Bernard, J.-P., Boulanger, F., Burton,
W. B., D\'{e}sert, F.-X., Hartmann, D. \ 1996, A\&A 308, L5  
\re
Rephaeli Y.\ 1995, ARA\&A 33, 541
\re
Rephaeli Y., Yankovitch, D. \ 1997, ApJ 481, L55 
\re
Rosati, P., Della Ceca, R., Burg R., Norman, C., Giacconi, R. \
1995, ApJ 445, L11
\re
Rosati, P., Della Ceca, R. \ 1997, in preparation
\re
Rosati, P., Della Ceca, R., Norman, C., Giacconi, R. \ 1997, ApJL submitted
\re 
Sasaki, S. \ 1994, PASJ 46, 427
\re
Suto, Y., Makishima, K., Ishisaki, Y., Ogasaka, Y. \ 1996,
ApJ 461, L33
\re
Sugiyama, N. \ 1995, ApJS 100, 281
\re
Sunyaev R.A., Zel'dovich Ya.B.\ 1972, Commts.\ Astrophys.\ Space Phys.\ 4, 
173
\re
Viana, P. T. P., Liddle, A. R. \ 1996, MNRAS 281, 323
\re
White, S. D. M., Efstathiou, G., Frenk, C. S. \ 1993, MNRAS 262, 1023

\label{last}

\begin{table}
\begin{center}
  Table~1.\hspace{4pt} CDM model parameters from the {\it ROSAT} X-ray
  \ns. \\ 
\end{center}
\vspace{6pt}
\begin{center}
\begin{tabular}{ccccccc}
\hline\hline\\[-6pt]
Model & $\Omega_0$ &  $\lambda_0$  
&  $h$ &   $\sigma_8$ & $\alpha$ & $\gamma$\\ 
[4pt]\hline \\[-6pt]
L03 & 0.3  & 0.7 & 0.7 & 1.04 & 3.4 & 1.2  \\
O045 & 0.45 & 0   & 0.7 & 0.83 & 3.4 & 1.2  \\
E1 & 1.0  & 0   & 0.5 & 0.56 & 3.4 & 1.2 \\
L03$\gamma$ & 0.3  & 0.7 & 0.7 & 0.90 & 3.4 & 1.5 \\
L01$\alpha$ & 0.1  & 0.9 & 0.7 & 1.47 & 2.7 & 1.2 \\
\hline
\end{tabular}
\end{center}
\end{table}

\end{document}